\documentclass[aps, showpacs, showkeys,nofootinbib,floatfix]{revtex4}

\usepackage{amssymb}
\usepackage{amsmath}
\usepackage{graphicx}

\begin{document}

\title{Reconstruction of a Deceleration Parameter from the Latest
Type Ia Supernovae Gold Dataset}

\author{Lixin Xu\footnote{Corresponding author}}
\email{lxxu@dl.cn}
\author{Chengwu Zhang}
\author{Baorong Chang}
\author{Hongya Liu}

\affiliation{School of Physics \& Optoelectronic Technology, Dalian
University of Technology, Dalian, 116024, P. R. China}

\begin{abstract}
In this paper, a parameterized deceleration parameter $q(z)= 1/2 -
a/(1 + z)^b$ is reconstructed from the latest type Ia supernovae
gold dataset. It is found out that the transition redshift from
decelerated expansion to accelerated expansion is at
$z_T=0.35^{+0.14}_{-0.07}$ with $1\sigma$ confidence level in this
parameterized deceleration parameter. And, the best fit values of
parameters in $1\sigma$ errors are $a=1.56^{+0.99}_{-0.55}$ and
$b=3.82^{+3.70}_{-2.27}$.
\end{abstract}



\hfill TP-DUT/2007-1

\maketitle

\section{Introduction}

Recent observations of High redshift Type Ia Supernova indicate that
our universe is undergoing accelerated expansion which is one of the
biggest challenges in present cosmological research, now
\cite{Riess98,Perlmuter99,Tonry03,Knop03,Barris03,Riess04}.
Meanwhile, this suggestion is strongly confirmed by the observations
from WMAP \cite{Bernardis00,Hanany00,Spergel03,Spergel06} and Large
Scale Structure survey \cite{Tegmark}. To understand the late-time
accelerated expansion of the universe, a large part of models are
proposed by assuming the existence of an extra energy component with
negative pressure and dominated at late time pushing the universe to
accelerated expansion. In principle, a natural candidate for dark
energy could be a small cosmological constant $\Lambda$ which has
the constant equation of state (EOS) $w_{\Lambda}=-1$. However,
there exist serious theoretical problems: fine tuning and
coincidence problems. To overcome the coincidence problem, the
dynamic dark energy models are proposed, such as quintessence
\cite{quintessence}, phantom \cite{phantom}, quintom \cite{quintom},
k-essence \cite{k-essence}, Chaplygin gas \cite{cha-gas},
holographic dark energy \cite{holo}, etc., as alternative
candidates.

Another approach to study the dark energy is by an almost
model-independent way, i.e., the parameterized equation state of
dark energy which is implemented by giving the concrete form of the
equation of state of dark energy directly, such as $w(z)=w_0+w_1 z$
\cite{Cooray99}, $w(z)=w_0+w_1\frac{z}{1+z}$
\cite{Cheallier00,Lider03},$w(z)=w_0+w_1\ln(1+z)$ \cite{Gerke02},
etc.. By this method, the evolution of dark dark energy with respect
to the redshift $z$ is explored, and it is found that the current
constraints favors a dynamical dark energy, though the cosmological
constant is not ruled out in $1\sigma$ region. But, the rapid
changed dark energy, $\left|\frac{\partial w}{\partial
z}\right|\ll1$, is ruled out \cite{Riess06}. Also, the dark energy
favors a quitom-like dark energy, i.e. crossing the cosmological
constant boundary. In all, it is an effective method to rule out the
dark energy models. As known, now the universe is dominated by dark
energy and is undergoing accelerated expansion. However, in the
past, the universe was dominated by dark matter and underwent a
decelerated epoch. So, inspired by this idea, the parameterized
decelerated parameter is present in almost model independent way by
giving a concrete form of decelerated parameters which is positive
in the past and changes into negative recently
\cite{Banerjee05,Xu06,Gong06}. Moreover, it is interesting and
important to know what is the transition time $z_T$ from decelerated
expansion to accelerated expansion. This is the main point of the
paper to be explored.

The structure of this paper is as follows. In section \ref{II}, a
parameterized decelerated parameter is constrained by latest $182$
Sne Ia data points compiled by Riess \cite{Riess06}. Section
\ref{III} is the conclusion.

\section{Reconstruction of Deceleration Parameter}\label{II}

We consider a flat FRW cosmological model containing dark matter and
dark energy with the metric
\begin{equation}
ds^2=-dt^2+a^2(t)dx^2.
\end{equation}
The Friedmann equation of the flat universe is written as
\begin{equation}
H^2=\frac{8\pi G}{3}\left(\rho_{m}+\rho_{de}\right),
\end{equation}
where, $H\equiv \dot{a}/a$ is the Hubble parameter, and its
derivative with respect to $t$ is
\begin{equation}
\dot{H}=\frac{\ddot{a}}{a}-\left(\frac{\dot{a}}{a}\right)^2,
\end{equation}
which combined with the definition of the deceleration parameter
\begin{equation}
q(t)=-\frac{\ddot{a}}{aH^2}
\end{equation}
gives
\begin{equation}
\dot{H}=-\left(1+q\right)H^2.\label{dotH}
\end{equation}
By using the relation $a_0/a=1+z$, the relation of $H$ and $q$, {\it
i.e.}, Eq. (\ref{dotH}) can be written in its integration form
\begin{equation}
H(z)=H_0\exp\left[\int_{0}^{z}\left[1+q(u)\right]d\ln(1+u)\right],
\end{equation}
where the subscript "$0$" denotes the current values of the
variables. If the function of $q(z)$ is given, the evolution of the
Hubble parameter is obtained. In this paper, we consider a
parameterized deceleration parameter \cite{Xu06},
\begin{equation}
q(z)=1/2-a/(1+z)^b,\label{q}
\end{equation}
where, $a$, $b$ are constants which can be determined from the
current observational constraints. From Eq. (\ref{q}), it can be
seen that at the limit of $z\rightarrow\infty$, the decelerated
parameter $q\rightarrow 1/2$ which is the value of decelerated
parameter at dark matter dominated epoch. And, the current value of
decelerated parameter is determined by $q_0=1/2-a$. In the Eq.
(\ref{q}) form of decelerated parameter, the Hubble parameter is
written in the form
\begin{equation}
H(z)=H_0(1 + z)^{3/2} \exp\left[a\left((1+z)^{-b}-1\right)/b\right]
\end{equation}
From the explicit expression of Hubble parameter, it can be seen
that this mechanism can also be tried as parametrization of Hubble
parameter.

Now, we can constrain the model by the supernovae observations. We
will use the latest released supernovae datasets to constrain the
parameterized deceleration parameter Eq. (\ref{q}). The Gold dataset
contains $182$ Sne Ia data \cite{Riess06} by discarding all Sne Ia
with $z<0.0233$ and all Sne Ia with quality='Silver'. The $182$
datasets points are used to constrain our model. Constraint from Sne
Ia can be obtained by fitting the distance modulus $\mu(z)$
\begin{equation}
\mu_{th}(z)=5\log_{10}(D_{L}(z))+\mathcal{M},
\end{equation}
where, $D_{L}(z)$ is the Hubble free luminosity distance $H_0
d_L(z)$ and
\begin{eqnarray}
d_L(z)&=&(1+z)\int_{0}^{z}\frac{dz^{\prime}}{H(z^{\prime})}\\
\mathcal{M}&=&M+5\log_{10}\left(\frac{H_{0}^{-1}}{Mpc}\right)+25
\nonumber\\
&=&M-5\log_{10}h+42.38,
\end{eqnarray}
where, $M$ is the absolute magnitude of the object (Sne Ia). With
Sne Ia datasets, the best fit values of parameters in dark energy
models can be determined by minimizing
\begin{equation}
\chi_{SneIa}^2(p_s)=\sum_{i=1}^{N}\frac{\left(\mu_{obs}(z_i)-\mu_{th}(z_i)\right)^2}{\sigma^2_{i}},
\end{equation}
where $N=182$ for Gold dataset, $\mu_{obs}(z_i)$s are the modulus
obtained from observations, $\sigma_{i}$ are the total uncertainty
of the Sne Ia data.

\begin{figure}[tbh]
\centering
\includegraphics[width=3.0in]{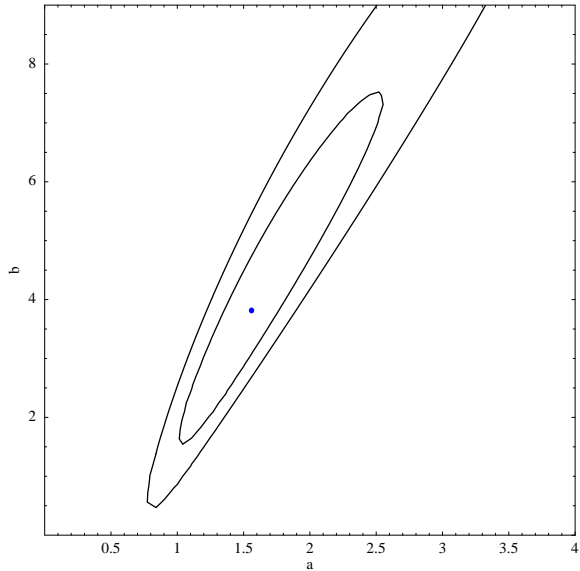}
\vspace*{8pt}
\caption{The contour plot of the parameter $a$ and $b$
with $1\sigma$ and $2\sigma$ confidence level.}\label{contour}
\end{figure}

Fitting the datasets from $182$ Gold Sne Ia, we obtain the minimum
$\chi^2_{min}$ with $182-2$ {\it dof} and the best fit parameters
$a$, $b$ and the transition time (redshift) $z_T$ in the
parameterized decelerated parameter, where $z_T$ defines the
transition redshift from decelerated expansion to accelerated
expansion, i.e., $q(z_T)=0$. The result is listed in the Table
\ref{results}. The $1\sigma$ and $2\sigma$ contours of parameters
$a-b$ are plotted in Fig. \ref{contour}.
\begin{table}[h]
{\begin{tabular}{@{}cccc@{}} \toprule
$\chi^2$ with $180$ {\it dof} & $a$& $b$ & $z_T$\\
\colrule
$156.66$ & $1.56^{+0.99}_{-0.55}$ & $3.82^{+3.70}_{-2.27}$ & $0.35^{+0.14}_{-0.07}$ \\
\botrule
\end{tabular}}
\caption{The Fitting Result from $182$ Gold Sne Ia, the best fit
parameters $a$, $b$ and $z_T$ with $1\sigma$ errors.}\label{results}
\end{table}

\begin{figure}[tbh]
\centering
\includegraphics[width=3.0in]{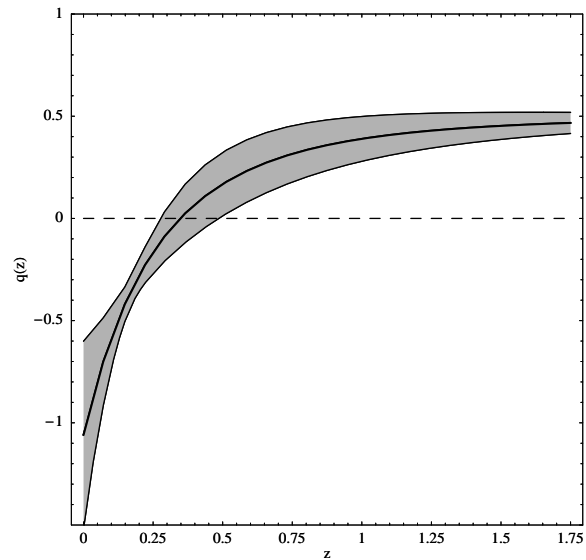}
\vspace*{8pt} \caption{The evolution of decelerated parameter with
respect to the redshift $z$. The center solid lines is plotted with
the best fit value, where the shadows denote the $1\sigma$
region.}\label{qz}
\end{figure}

In our reconstruction, measurement errors are considered by using
the well-known error propagation equation for any $y(x_1, x_2, ...,
x_n)$,
\begin{equation}
\sigma^{2}(y)=\sum^{n}_{i=1}\left(\frac{\partial y}{\partial
x_i}\right)_{x=\bar{x}}^2
Cov(x_i,x_i)+2\sum^{n}_{i=1}\sum^{n}_{j=i+1}\left(\frac{\partial
y}{\partial x_i}\right)_{x=\bar{x}}\left(\frac{\partial y}{\partial
x_j}\right)_{x=\bar{x}} Cov(x_i,x_j)
\end{equation}
is used extensively (see Ref. \cite{Alam} for instance). For Ansatz
Eq. (\ref{q}), we obtain errors of the parameterized decelerated
parameter. The evolution of the decelerated parameters $q(z)$ with
$1\sigma$ error are plotted in Fig. \ref{qz}.

\section{Conclusion}\label{III}

In this paper, by an almost model-independent way, we have used a
parameterized decelerated parameter to obtain the transition time or
redshift $z_T$ from decelerated expansion to accelerated expansion.
It is found out that the best fit transition redshift $z_T$ is about
$z_T=0.35^{+0.14}_{-0.07}$ with $1\sigma$ error in this
parameterized equation which is compatible with the result of Ref.
\cite{Riess06}. Though, we also can derive the transition redshift
from a giving equation of state of dark energy and an concrete dark
energy models, they are much model dependent. So, we advocate the
almost model-independent way to test and rule out some existent dark
energy models.

\section*{Acknowledgments}
L. Xu is supported by DUT (3005-893321) and NSF (10647110). H. Liu
is supported by NSF (10573003) and NBRP (2003CB716300) of P.R.
China.

\end{document}